\documentclass[prd,eqsecnum,twocolumn,showpacs,amsmath,amssymb]{revtex4}

\usepackage{graphicx}

\usepackage{bm}

\newcommand{\beq}{\begin{equation}}
\newcommand{\eeq}{\end{equation}}
\newcommand{\beqs}{\begin{eqnarray}}
\newcommand{\eeqs}{\end{eqnarray}}

\newcommand{\drawsquare}[2]{\hbox{%
\rule{#2pt}{#1pt}\hskip-#2pt
\rule{#1pt}{#2pt}\hskip-#1pt
\rule[#1pt]{#1pt}{#2pt}}\rule[#1pt]{#2pt}{#2pt}\hskip-#2pt
\rule{#2pt}{#1pt}}
\newcommand{\fund}{\raisebox{-.5pt}{\drawsquare{6.5}{0.4}}}

\begin{document}

\title{Some Results on Vector and Tensor Meson Mixing in a Generalized
QCD-like Theory}

\author{Hai-Yang Cheng$^{a,b}$}

\author{Robert Shrock$^b$}

\affiliation{(a)
Institute of Physics, Academia Sinica, Taipei, Taiwan 115, Republic of China}

\affiliation{(b)
C.N. Yang Institute for Theoretical Physics, Stony Brook University,
Stony Brook, NY 11794}

\begin{abstract}

We present results for the mixing of quark flavor eigenstates to form low-lying
vector and tensor meson mass eigenstates in a generalized QCD-like theory with
$\ell$ massless or light, degenerate quarks and one quark of substantial mass,
$m_Q$.  We show that as $m_Q/\Lambda$ gets large, where $\Lambda$ is the
confinement scale, the angle governing this mixing approaches
$\arctan(1/\sqrt{\ell})$.  We give a generalization of the Gell-Mann Okubo mass
relation for this theory.  A comparison of the $\ell=2$ results with data on
$\omega$-$\phi$ and $f_2(1275)$-$f_2'(1525)$ mixing is included.

\end{abstract}

\pacs{}

\maketitle

\section{Introduction}
\label{intro}

The mixing of the flavor-SU(3) singlet and octet states of vector and tensor
mesons to form mass eigenstates is of fundamental importance in hadronic
physics. In the case of the vector mesons, with $J^{PC}=1^{--}$, the
flavor-SU(3) singlet, $\omega_0$ and octet, $\omega_8$, mix through an angle
$\theta_V \simeq 39^\circ$ \cite{pdg}, close to the value that would lead to
the complete decoupling of the light $u$ and $d$ quarks from the heavier $s$
quark in the resultant mass eigenstates $\omega$ and $\phi$. Throughout the
years there have been many studies of this mixing \cite{earlyvm}-\cite{vmlist}.
In a modern context, some insight into this comes from the Appelquist-Carazzone
decoupling theorem \cite{ac}, according to which, in a vectorial theory, as the
mass of a particle gets large compared with a relevant scale - here the scale
of confinement and spontaneous chiral symmetry breaking (S$\chi$SB) in quantum
chromodynamics (QCD), $\Lambda_{QCD} \simeq 300$ MeV - one can integrate this
particle out and define a low-energy effective field theory applicable below
this scale.  Evidently, even though $m_s$ is not $>> \Lambda_{QCD}$, there is
still a nearly complete decoupling, so that the physical $\omega$ is mostly
comprised of the isospin-singlet combination $(u \bar u + d \bar d)/\sqrt{2}$,
while $\phi$ is mostly an $s \bar s$ state.  The $s \bar s$ constituency of the
$\phi$ is inferred from the fact that $\phi$ has a dominant ($> 80 \%$)
branching ratio to $K^+ K^-, \ K_L^0 K_S^0$, despite the fact that there is
very little phase space for these decay modes, much less than the phase space
available for $\phi \to \rho \pi$ and $\phi \to 3\pi$ decay modes. This
motivated the Okubo-Zweig-Iizuka (OZI) rule \cite{ozi}.  A similar situation of
near-ideal mixing occurs for the $J^{PC}=2^{++}$ tensor mesons
$f_2(1275)$-$f_2'(1525)$, and can also be understood in terms of approximate
decoupling of the light $u \bar u + d \bar d$ state from the heavier $s \bar s$
state.

In this paper we discuss the mixing of self-conjugate mesons in a generalized
QCD-like theory with emphasis on the role of decoupling. Specifically, we
analyze the mixing of quark flavor eigenstates to form the lowest-lying mass
eigenstates of vector and tensor mesons in a vectorial, asymptotically free,
confining SU($N_c$) gauge theory, where $N_c \ge 3$. We denote the scale of
confinement and spontaneous chiral symmetry breaking in this theory as
$\Lambda$. The fermion content consists of $N_f$ fermions transforming
according to the fundamental representation of SU($N_c$), i.e., the quarks, of
which
\beq
\ell=N_f-1
\label{ell}
\eeq
are massless (or of small, equal mass $m_q << \Lambda$) and
one has a larger mass, $m_Q > m_q$.  We presume that electroweak interactions
have been turned off, so that only color SU($N_c$) interactions are present.
We define the ratio
\beq
r_Q = \frac{m_Q}{\Lambda} \ .
\label{rq}
\eeq
We first derive the decoupling value of the mixing angle and then discuss how
the mixing angle varies as a function of $m_Q/\Lambda$.  For the special case
$N_c=3$ and $\ell=2$, the framework for our analysis is a rough approximation
to QCD with the heavy quarks, $c$, $b$, and $t$ integrated out, since, the
hard masses $m_{u,d} << \Lambda_{QCD}$ \cite{hard,mud,cpv}, $m_s \sim
100$ MeV is smaller than, but comparable to $\Lambda_{QCD}$, and electroweak
interactions are a small perturbation. Accordingly, we remark on the connection
of our results with the mixing of the vector mesons $\omega$ and
$\phi$, and the tensor mesons $f_2(1275)$ and $f_2'(1525)$.  We also
derive a generalization of the Gell-Mann Okubo mass relation for this theory

In general, when analyzing flavor-singlet states, it is not just the quark
flavor-eigenstates that one must include in the initial basis of states, but
also pure gluonic states with the same $J^{PC}$.  However, the masses of the
lowest-lying glueballs with the relevant $J^{PC}$ values, as determined by
lattice measurements, are well above the masses of the relevant quark-antiquark
states.  Specifically, these lattice measurements yield (i) a value of $\sim 4$
GeV for the lowest-lying glueball with $J^{PC}=1^{--}$, considerably greater
than the masses of the $\omega$ and $\phi$ (and also above the mass of the
$J/\psi$ and $\psi'$); and (ii) a mass $\sim 2.5$ GeV for the lowest-lying
glueball with $J^{PC}=2^{++}$, which is substantially larger than the masses of
the $f_2(1275)$ and $f_2'(1525)$ tensor mesons \cite{lgtglue}. Therefore, it
should be a reasonable approximation to neglect mixing of quark-antiquark
flavor eigenstates with these gluonic states to form the lowest-lying meson
mass eigenstates.  In contrast, as is well known, for the $J^{PC}=0^{++}$
scalar mesons in the mass interval 1-2 GeV, the mixing of quark-antiquark and
glue states is important.

\section{Meson Mixing in a Generalized QCD-Like Theory}
\label{generalmixing}

In this section we discuss the mixing of vector and tensor meson flavor
eigenstates to form the lowest-lying mass eigenstates in a generalized QCD-like
SU($N_c$) gauge theory.  We denote the $\ell$ quarks with zero mass or a common
light mass $m_q << \Lambda$ as $q_i$, $i=1,...,\ell$, and the one with the
larger mass, $m_Q$, as $q_{_{N_f}} \equiv Q$.  We will usually assume that
$\ell \ge 2$, since for $\ell=1$ the theory has no nontrivial isospin
symmetry. It is assumed that $N_f$ is sufficiently small that the theory is not
only asymptotically free, but confines and spontaneously breaks chiral symmetry
\cite{su3nfc}. If the $\ell$ quarks are massless, this theory has a
(nonanomalous) global flavor symmetry $G_{f,\ell}={\rm SU}(\ell)_L \otimes {\rm
SU}(\ell)_R \otimes {\rm U}(1)_V$.  (If the $\ell$ quarks have nonzero masses
$m_q << \Lambda$, then the chiral part of this symmetry is approximate.) The
condensates of the massless quarks, $\langle \bar q_i q_i \rangle$, where
$\ell=1,...,\ell$, preserve the U(1)$_V$ and, since we neglect the small
electromagnetic interactions \cite{vacalign}, they are all equal. Hence, they
spontaneously break the chiral part of $G_{f,\ell}$ to its diagonal, vectorial
(isospin) subgroup, ${\rm SU}(\ell)_V$. This residual ${\rm SU}(\ell)_V$
isospin symmetry (as well as the U(1)$_V$ symmetry) is also present if the
$\ell$ quarks have nonzero degenerate masses.  The subscript $V$ will
henceforth be dropped when the meaning is clear.  In the absence of an
underlying dynamical mechanism to render the $\ell$ light quarks degenerate,
one may regard the condition of degeneracy at a nonzero value of mass as
artificial; the reason that we include it here is that in our application to
QCD, we will make contact with quark model results, and these often take the
$u$ and $d$ quarks to have a common effective mass $m_q$.

Now we treat $m_Q$ as a variable, and start with it equal to its lower limit,
$m_Q=m_q$.  In this limit, the global chiral symmetry is ${\rm SU}(N_f)_L
\otimes {\rm SU}(N_f)_R$, which is broken to ${\rm SU}(N_f)_V$.  The natural
basis of states for $m_Q=m_q$ is the basis for ${\rm SU}(N_f)_V$, with the
(diagonal) generators of the $(N_f-1)$-dimensional Cartan algebra of ${\rm
SU}(N_f)$ given by the $N_f \times N_f$ matrices
\beq
T_a \ , \quad a=k^2-1 \ {\rm for} \ k=1,..., N_f \ .
\label{tadiagonal}
\eeq
For example, $T_3 = (1/2){\rm diag}(1,-1,0,...,0)$,
$T_8 = (1/\sqrt{12} \, ) {\rm diag}(1,1,-2,0...,0)$, etc.
The corresponding neutral vector meson states are
\beq
|V_a \rangle = \sqrt{2} \,
\Big | \sum_{i,j=1}^{N_f} q_i (T^a)_{ij} \bar q_j \Big \rangle \ .
\label{va}
\eeq
Explicitly, with $\ell=N_f-1$,
\beq
|V_a \rangle \equiv |V_{N_f^2-1}\rangle =
\Big | \frac{(\sum_{i=1}^\ell q_i \bar q_i) -\ell Q \bar Q}
{\sqrt{\ell(\ell+1)}} \Big \rangle \ .
\label{vaexplicit}
\eeq
We also define the flavor-${\rm SU}(n)$-singlet meson state
\beq
|(V_0)_n \rangle = \frac{1}{\sqrt{n}} \,
\Big | \sum_{i=1}^{n} q_i \bar q_i \Big \rangle \ ,
\label{v0k}
\eeq
which will be relevant for $n=\ell$ and $n=N_f$.

We now increase $m_Q$ from $m_q$ to larger values.  This breaks the ${\rm
SU}(N_f)$ isospin symmetry to ${\rm SU}(\ell)$.  For $m_q = 0$, the size of
this breaking is determined by the ratio $r_Q$.  For the case of $m_q \ne 0$,
there are two relevant measures of this breaking, $r_Q$ and $m_Q/m_q$, but if
$m_q << \Lambda$, then one expects that the effect is primarily measured by
$r_Q$.  Of the $N_f^2$ vector mesons, $\ell^2$ transform according to the
irreducible representations of ${\rm SU}(\ell)$ that occur in the
Clebsch-Gordan decomposition ${\fund}_\ell \times \overline{\fund}_\ell =
(adj)_\ell + 1_\ell$, where ${\fund}_\ell$, $adj_\ell$, and $1_\ell$ denote the
fundamental, adjoint and singlet of ${\rm SU}(\ell)$.  In the ${\rm SU}(N_f)$
basis, the mixing of states must respect the ${\rm SU}(\ell)$ symmetry. This
only allows the mixing of the states corresponding to flavor-diagonal bilinears
whose ${\rm SU}(\ell)$ parts are proportional to $(V_0)_\ell$, i.e.,
SU($\ell$)-singlets, and, in the quark-antiquark sector, there are only two
such states, namely $|(V_0)_{N_f}\rangle$ and $|V_{N_f^2-1}\rangle$.
Pure gluonic states are also SU($\ell$)-singlets and can, in
general, mix with these SU($\ell$)-singlet quark-antiquark flavor
eigenstates. However, as noted above, the masses of these pure gluonic states
are sufficiently large that it should be reasonable to neglect them in our
analysis. Accordingly, we focus on the mixing of the
$|(V_0)_{N_f}\rangle$ and $|V_{N_f^2-1}\rangle$ states.
We label the two resultant mass eigenstates as $|V_L \rangle$ and
$|V_H\rangle$, with masses $m_L$ and $m_H$, respectively, where the subscripts
$L$ and $H$ stand for lighter and heavier.

We next discuss this diagonalization in greater detail. We analyze the
mass-squared matrix $M^2$ in the basis of flavor eigenstates $(|(V_0)_{N_f}
\rangle, \, |V_{N_f^2-1}\rangle)$ \cite{masssquared}. In general, one should
take into account the fact that the particles involved are unstable, with
strong decay widths that are substantial fractions of their masses.  However,
one may obtain an approximate description of the physics by neglecting the
widths. In this case we can write $M^2$ as the real, symmetric matrix
\beq
M^2 = \left( \begin{array}{cc}
    m_1^2     & \delta \\
    \delta    & m_a^2    \end{array} \right ) \ ,
\label{msq}
\eeq
where the subscripts 1 and $a$ refer to flavor ${\rm SU}(N_f)$ singlet and
\underline{a}djoint.  One can work backward from the observed masses and mixing
angle to determine $\delta$.  Note that $|\delta|$ is bounded as
to
\beq
|\delta| < m_1 m_a \ ,
\label{deltabound}
\eeq
since a value of $|\delta|$ as large as $m_1m_a$ would
have the consequence that the determinant ${\rm det}(M^2)=0$ and hence that the
lighter vector meson mass, $m_L$, would vanish.  Indeed, since
$m_L$ is of order $2\Lambda_{QCD}$, it follows that $|\delta|$ is considerably
smaller than this upper bound, and the explicit results satisfy this.

The matrix $M^2$ is diagonalized according to
\beq
R(\theta_V) M^2 R(\theta_V)^{-1} = M^2_{diag.}
\label{rmrinv}
\eeq
where
\beq
M^2_{diag.} = \left( \begin{array}{cc}
     m_L^2 & 0   \\
       0   & m_H^2  \end{array} \right )
\label{msqdiag}
\eeq
with the eigenvalues of $M^2$ given by
\beq
m_{H,L}^2 = \frac{1}{2}\bigg [ m_a^2+m_1^2 \pm \sqrt{(m_a^2-m_1^2)^2
+ 4\delta^2} \ \bigg ] \ .
\label{mhl}
\eeq
In Eq. (\ref{rmrinv}), the orthogonal rotation matrix $R(\theta_V)$
is
\beq
R(\theta_V) = \left( \begin{array}{cc}
    \cos\theta_V & \sin\theta_V   \\
   -\sin\theta_V & \cos\theta_V  \end{array} \right ) \ ,
\label{rmatrix}
\eeq
where
\beq
\tan(2\theta_V) = \frac{2\delta}{m_1^2-m_a^2} \ .
\label{tan2theta}
\eeq
Note that this equation only determines $\theta_V$ up to a shift by $\pm \pi$.
The masses $m_L$ and $m_H$ satisfy the trace and determinant
relations
\beq
m_L^2+m_H^2 =  {\rm Tr}(M^2) = m_1^2+m_a^2
\label{tracerel}
\eeq
and
\beq
m_L^2 \, m_H^2 = {\rm det}(M^2) = m_1^2m_a^2-\delta^2 \ .
\label{detrel}
\eeq
The flavor eigenstates are mapped to mass eigenstates according to
\beq
 \left( \begin{array}{c}
    |V_L\rangle \\
    |V_H \rangle \end{array} \right ) =
R(\theta_V) \left( \begin{array}{c}
    |(V_0)_{N_f} \rangle \\
    |V_{N_f^2-1} \rangle \end{array} \right ) \ .
\label{generalmapping}
\eeq

We first comment on the limit where $m_Q=m_q$, so that the SU($\ell$) flavor
(generalized isospin) symmetry is increased to an ${\rm SU}(\ell+1)={\rm
SU}(N_f)$ flavor symmetry.  First, the SU($N_f$) symmetry implies that
$\delta=0$, since the (1,2) and (2,1) elements of $M^2$ connect two different
representations of SU($N_f$) (the singlet and adjoint). The SU($N_f$) flavor
symmetry requires that all of the members of a given representation of
SU($N_f$) must be degenerate.  Although it does not, by itself, imply that
members of different SU($N_f$) representations are degenerate, this is observed
empirically to be approximately true in QCD.  For example, the $\rho$ and
$\omega$, which, to very good accuracy, are adjoint and singlet states with
respect to isospin SU(2), have nearly equal masses, with a fractional
difference $(m_\omega-m_\rho)/[(1/2)(m_\omega+m_\rho)] \simeq 0.01$.  In our
generalized context, we may thus anticipate that the SU($N_f$)-singlet and
adjoint states $|(V_0)_{N_f} \rangle$ and $|V_{N_f^2-1} \rangle$ are
approximately degenerate.  Indeed, let us consider the case where these states
are exactly degenerate, i.e., $m_1=m_a$. In this case, any value of $\theta_V$
in Eq. (\ref{generalmapping}) yields a corresponding set of degenerate mass
eigenstates. This is reflected in the property that the right-hand side of
Eq. (\ref{tan2theta}) is proportional to the indeterminate ratio 0/0.

One may also formally set $\delta = 0$ without setting $m_q = m_Q$ or
$m_1=m_a$. The $|(V_0)_{N_f} \rangle$ and $|V_{N_f^2-1}\rangle$ are then mass
eigenstates, with masses $m((V_0)_{N_f}) = m_1$ and $m(V_{N_f^2-1}) = m_a$.
The $|(V_0)_{N_f} \rangle$ state has a probability $1/N_f$ of being in one of
the $|q_i \bar q_i\rangle$ states, for $i=1,...,\ell$, and hence a probability
$\ell/N_f = \ell/(\ell+1)$ of being in some $|q_i \bar q_i \rangle$ state, and
a probability $1/N_f$ of being in the $|Q \bar Q\rangle$ state.  In contrast,
the $|V_{N_f^2-1} \rangle$ state has a probability $1/[\ell(\ell+1)]$ of being
in one of the $|q_i \bar q_i\rangle$ states, for a total probability of $1/N_f$
of being in some $|q_i \bar q_i \rangle$ state, and a probability
$\ell/(\ell+1)=\ell/N_f$ of being in the $|Q \bar Q \rangle$ state.  The pair
$Q \bar Q$ comprised of the heavier quark, thus enters the wavefunction for the
$|V_{N_f^2-1} \rangle$ state with a probability $\ell/N_f$ which, for the $\ell
\ge 2$ case under consideration here, is greater than the probability $1/N_f$
with which it enters into the wavefunction for the $|(V_0)_{N_f}\rangle$ state.
If these considerations formed a complete analysis, they would imply that for
$\ell \ge 2$, and $m_Q > m_q$, the $|V_{N_f^2-1} \rangle$ state would be
heavier than the $|(V_0)_{N_f}\rangle$ state, i.e, $m_a > m_1$.  This argument
is borne out by the physical mass ordering in Eq. (\ref{msqphysical}) below
(and for tensor mesons, in Eq. (\ref{msqtensor})).  We also mention an approach
in which these mesons are modelled as dynamical fields in an effective
Lagrangian, with both kinetic and mass terms. Although one cannot do
perturbative calculations reliably because of the strongly coupled nature of
the physics, one may note that there would be a particular kind of correction
to the propagators of the $|(V_0)_{N_f}\rangle$ state, in which it makes a
transition to a glueball state of the same $J^{PC}$ and back again \cite{ann}.
Insofar as one can use perturbative degrees of freedom to describe this
transition, for a $J^{PC}=1^{--}$ vector meson it would involve a graph in
which the quark-antiquark pair annihilates into three or, more generally, an
odd number of gluons (from $C$ invariance) and then these create the
quark-antiquark final state $|(V_0)_{N_f}\rangle$. For a $J^{PC}=2^{++}$ tensor
meson, the intermediate gluonic state would consist of an even number of
gluons.  Since our QCD-like theory is asymptotically free, as in QCD, the
larger $m_Q$ is, the smaller the gauge couplings multiplying the vertices
involving gluon emission from the $Q$ line \cite{ann}.  This type of propagator
correction would not occur for the $|V_{N_f^2-1} \rangle$, since it is not a
flavor-singlet.  As mentioned above, the lowest-lying glueball state with
$J^{PC}=1^{--}$ has a mass considerably larger than the $\omega$ and $\phi$
masses, so it seems reasonable to neglect this propagator correction.  With our
ordering of the flavor eigenbasis as $(|(V_0)_{N_f}\rangle, \,
|V_{N_f^2-1}\rangle)$ and the mass eigenbasis as $(|V_L\rangle, \, |V_H
\rangle)$ and with $\delta=0$, it follows that if $m_a > m_1$, then
$\theta_V=0$ and if $m_1 > m_a$, then $\theta=\pi/2$. (In each case, the other
value of $\theta$ is excluded because it would map the lighter flavor state to
$|V_H\rangle$ and the heavier flavor state to $|V_L\rangle$.)

We now proceed to the general case of nonzero $\delta$. First, we observe that
from Eq. (\ref{tan2theta}) it follows that if $m_a > m_1$ ($m_1 < m_a$), then
the sign of $\tan(2\theta_V)$ is opposite to (the same as) the sign of
$\delta$.  The value of $\theta_V$ that produces exact decoupling ($dec.$) of
the $\ell$ massless quarks $q_i$, with $i=1,...,\ell$ from the one heavier
quark, $Q$, is determined by the condition that the mass eigenstates are
$|V_L\rangle = |(V_0)_{\ell}\rangle$ and $|V_H\rangle = -|V_{Q \bar Q}\rangle$
(where the minus sign is included for convenience and has no physical
significance).  Imposing this decoupling condition on the transformation
(\ref{generalmapping}), we derive the decoupling value of the mixing angle,
\beq
\theta_{V,dec.} = \arctan \Big ( \frac{1}{\sqrt{\ell}} \Big ) \ .
\label{thetadec}
\eeq

In this situation of exact decoupling, the ${\rm SU}(N_f)_V$ eigenstates
$|(V_0)_{N_f}\rangle$ and $|V_{N_f^2-1}\rangle$ are mapped to the mass
eigenstates $|V_L\rangle = |(V_0)_\ell \rangle$ and $|V_H\rangle = -|V_{Q \bar
Q} \rangle$ according to
\beq
 \left( \begin{array}{c}
    |V_L \rangle \\
    |V_H \rangle \end{array} \right ) =
 \left( \begin{array}{c}
    |(V_0)_\ell\rangle \\
    -|V_{Q \bar Q} \rangle \end{array} \right ) =
R(\theta_{V,dec.}) \left( \begin{array}{c}
    |(V_0)_{N_f} \rangle \\
    |V_{N_f^2-1} \rangle \end{array} \right ) \ .
\label{mapping_dec}
\eeq
For the decoupling value $\theta_V=\theta_{V,dec.}$, we have
\beq
R(\theta_{V,dec.}) = \left( \begin{array}{cc}
    \sqrt{\frac{\ell}{\ell+1}} & \frac{1}{\sqrt{\ell+1}}   \\
   -\frac{1}{\sqrt{\ell+1}}    &  \sqrt{\frac{\ell}{\ell+1}}
 \end{array} \right ) \ .
\label{rmatrixdec}
\eeq
In the decoupling limit, the mass-squared matrix $M^2$ in the
basis of flavor eigenstates $|(V_0)_{N_f} \rangle$ and $|V_{N_f^2-1}\rangle$
can be expressed as
\beqs
(M^2)_{dec.} & = & R(\theta_{dec.})^{-1} M^2_{diag.} R(\theta_{dec.}) \cr\cr
    & = & \left( \begin{array}{cc}
    \frac{\ell m_L^2 + m_H^2}{\ell+1} &
    -\frac{\sqrt{\ell} \, (m_H^2-m_L^2)}{\ell+1} \\
    -\frac{\sqrt{\ell} \, (m_H^2-m_L^2)}{\ell+1} &
    \frac{m_L^2 + \ell m_H^2}{\ell+1} \end{array} \right ) \ .
\cr\cr
& &
\label{msq_dec}
\eeqs
Equating the (1,1) and (2,2) elements of $(M^2)_{dec.}$ in Eq.
(\ref{msq_dec}) to the corresponding elements of the general expression for
$M^2$ in Eq. (\ref{msq}), we obtain relations for $m_L^2$ and $m_H^2$ in terms
of $m_1^2$ and $m_a^2$ for this decoupling case (applicable for the nontrivial
range $\ell \ge 2$), namely
\beq
m_L^2 = \frac{\ell m_1^2-m_a^2}{\ell-1}
\label{mLsq_dec}
\eeq
and
\beq
m_H^2 = \frac{\ell m_a^2-m_1^2}{\ell-1} \ .
\label{mHsq_dec}
\eeq
For given values of $m_1$ and $m_a$ and for the nontrivial range $\ell \ge 2$
of interest here, one can determine the value of $\delta$ that
leads to decoupling by setting $\theta_V = \theta_{V,dec.}$, substituting this
into Eq. (\ref{tan2theta}).  This yields the result
\beq
\delta_{dec.} = -\Bigg ( \frac{\sqrt{\ell}}{\ell-1} \Bigg ) (m_a^2-m_1^2) \ .
\label{deltadec}
\eeq
Combining the general inequality (\ref{deltabound}) and Eq. (\ref{deltadec}),
we derive an upper bound on $|\delta_{dec.}|$, namely
\beq
\Bigg ( \frac{\sqrt{\ell}}{\ell-1} \Bigg ) (m_a^2-m_1^2) < m_1 m_a \ .
\label{deltadecbound}
\eeq
As with the general bound (\ref{deltabound}), one actually expects the
left-hand side of this bound to be substantially smaller than the right-hand
side, since otherwise, the diagonalization of $M^2$ in the decoupling limit
would lead to the smaller squared-mass eigenvalue $m_L^2$ being much less than
the typical QCD scale, $\Lambda_{QCD}^2$, which is unphysical, since the
corresponding hadron is not a (pseudo)-Nambu-Goldstone boson.

The off-diagonal element of (\ref{msq_dec}) gives an equivalent expression for
$\delta_{dec.}$, viz., $\delta_{dec.} =
-\sqrt{\ell} \, (m_H^2-m_L^2)/(\ell+1)$.  Setting these two expressions equal,
we derive a relation that applies if there is exact decoupling:
\beq
\frac{m_H^2-m_L^2}{m_a^2-m_1^2} = \frac{\ell+1}{\ell-1} \ .
\label{massrel}
\eeq

As $m_Q$ increases to values far above $\Lambda_{QCD}$, i.e., for $r_Q >> 1$,
one can define an effective low-energy field theory with the quark $Q$
integrated out \cite{ac} and therefore the mixing of the quark-antiquark flavor
eigenstates must lead to decoupling of this heavy quark. Imposing this
decoupling condition on the transformation (\ref{generalmapping}), we deduce
that
\beq
{\rm If} \ r_Q \to \infty, \quad {\rm then} \quad \theta_V \to
\theta_{V,dec.} = \arctan \bigg ( \frac{1}{\sqrt{\ell}} \bigg ) \ .
\label{theta_dec}
\eeq
Our explicit calculations are in accord with this general result, since in
this limit, $m_H/m_L \to \infty$, so that
\beq
M^2 \to  \frac{m_H^2}{\ell+1} \, \left( \begin{array}{cc}
  1       & -\sqrt{\ell} \\
  -\sqrt{\ell} &    \ell  \end{array} \right ) \ .
\label{msq_asymp}
\eeq
The diagonalization of this limiting expression for $M^2$ yields $\tan(2\theta)
= 2\sqrt{\ell}/(\ell-1)$, i.e., $\tan\theta=\tan\theta_{dec.} = 1/\sqrt{\ell}$.

In general, as discussed above, the mixing involves not just quark-antiquark
flavor-eigenstates, but also gluonic states of the same $J^{PC}$, to form the
mass eigenstates.  In this more general framework one can describe this mixing
via the orthogonal $3 \times 3$ transformation that maps the quark-antiquark
flavor eigenstates $|(V_0)_{N_f} \rangle$, $|V_{N_f^2-1} \rangle$, and the glue
state $|G\rangle$ with the same $J^{PC}$ to the three resultant lowest-lying
mass eigenstates, denoted $|V_L\rangle$, $|V_H\rangle$, and $|V_G\rangle$,
\beq
\left( \begin{array}{c}
  |V_L\rangle \\
  |V_H \rangle \\
  |V_G \rangle \end{array} \right ) =
R_{gen.} \left( \begin{array}{c}
  |(V_0)_{N_f} \rangle \\
  |V_{N_f^2-1} \rangle \\
  |G            \rangle \end{array} \right ) \ .
\label{3x3mapping}
\eeq
As an element of the orthogonal group O(3), $R_{gen.}$ (with subscript
$gen.$ for ``general'') depends on three Euler angles.  We write
\begin{widetext}
  \beq
  R_{gen.} = \left ( \begin{array}{ccc}
    \cos(\theta_{qG}) &   0   & \sin(\theta_{qG}) \\
    0          &   1   &      0           \\
    -\sin(\theta_{qG}) &   0   & \cos(\theta_{qG}) \end{array} \right )
  \left ( \begin{array}{ccc}
    1     &       0          &      0          \\
    0     & \cos(\theta_{QG})  & \sin(\theta_{QG}) \\
    0     & -\sin(\theta_{QG}) & \cos(\theta_{QG}) \end{array} \right )
  \left ( \begin{array}{ccc}
    \cos(\theta_V)   & \sin(\theta_V)    &  0 \\
    -\sin(\theta_V)  & \cos(\theta_V)    &  0 \\
    0         &      0          &  1 \end{array} \right ) \ .
  \label{rgen}
  \eeq
\end{widetext}
The decoupling theorem implies that as $r_Q \to \infty$ and the heavy quark,
$Q$, is integrated out, the resultant low-energy effective field theory has the
property that there is no mixing of a $|Q \bar Q\rangle$ state with either $|q
\bar q\rangle$ or $|G\rangle$.  Hence, reading from right to left in
Eq. (\ref{rgen}), (i) in the first transformation, the angle $\theta_V$ takes
the value given by Eq. (\ref{theta_dec}), so that the vector of states
$(|(V_0)_{N_f} \rangle, \, |V_{N_f^2-1} \rangle, \, |G \rangle \, )^T$ is
mapped to $(|(V_0)_\ell\rangle, \, -|V_{Q \bar Q} \rangle, \, |G\rangle \,
)^T$; and (ii) in the second transformation,
\beq
{\rm If} \ r_Q \to \infty, \quad {\rm then} \quad \theta_{QG} \to 0 \ .
\label{theta_QG_dec}
\eeq
The third (i.e., left-most) transformation describes the mixing of the
light-quark $|q \bar q\rangle$ and $|G\rangle$ states.  This is not directly
relevant to our present analysis on decoupling of the quark $Q$ as $r_Q \to
\infty$, but we note that, as discussed above, this mixing is small because of
the considerable separation in mass between the gluonic states and the $|q \bar
q\rangle$ states with the same $J^{PC}$ values.

 The mixings of $|(V_0)_{N_f} \rangle$, $|V_{N_f^2-1} \rangle$, and
$|G\rangle$ depend in a complicated way on $r_Q$.  As $m_Q$ increases above
$m_q$, the angle $\theta_V$ moves upward from 0 or downward from $\pi/2$,
depending on its initial value for $m_Q = m_q$.  On the basis of decoupling
arguments, the variation of $\theta_V$ as a function of $m_Q/\Lambda$ should be
monotonic.  The decoupling result (\ref{theta_dec}) holds concerning the mixing
of the $|Q \bar Q\rangle$ state with the lowest-lying $|q \bar q\rangle$ state
of the same $J^{PC}$.

  As $m_Q$ increases sufficiently, the mass of a $|Q \bar Q\rangle$ state will
become equal to the mass of the $|G\rangle$ state with the same $J^{PC}$, and
there will be strong mixing between these two states, which will be reflected
in $\theta_{QG}$.  Eventually, as $m_Q$ increases far above the confinement
scale $\Lambda$, the mixing angles $\theta_V$ and $\theta_{QG}$ will approach
their respective asymptotic values, (\ref{theta_dec}) and (\ref{theta_QG_dec}).
Decoupling arguments also imply that for $m_Q$ above the mass of the lightest
gluonic state, the approach of $\theta_{QV}$ to zero should also be monotonic.
These statements apply for the mixings with the lowest-lying mass states,
$(|V_L\rangle, \, |V_M\rangle, \, |V_H\rangle \, )^T$, in
Eq. (\ref{3x3mapping}).  It should be noted that as $m_Q$ increases, the mass
of a $|Q \bar Q\rangle$ state will ascend through values equal to the masses of
various higher radial and orbital excitations of light-quark $|q \bar q\rangle$
states, and with excited gluonic states, with the same $J^{PC}$, so that one
would have to take account of the mixing with these excited states also.  Let
us recall the standard spectroscopic notation for fermion-antifermion bound
states, $n \, {}^{2S+1}L_J$, where $n$ denotes the radial quantum number, and
$\vec S$, $\vec L$, and $\vec J$ denote the total spin, orbital angular
momentum, and total angular momentum of the state. As an example for vector
mesons, as $m_Q$ increases sufficiently, the mass of a $|Q \bar Q\rangle$ state
would sequentially pass through the values of the masses of the $n \, {}^3 S_1$
light-quark $|q \bar q\rangle$ states with $n \ge 2$, as well as the $1 \, {}^3
D_1$ $|q \bar q\rangle$ state, resulting in significant mixing with these
states.  However, in general, for higher masses, the widths of these states
become larger, which would have the effect of reducing the mixing.

\section{Generalized Gell-Mann Okubo Mass Relation}
\label{ggmorsection}

In this section we derive the generalization of the Gell-Mann Okubo (GMO) mass
relation for vector and tensor mesons in our SU($N_c$) QCD-like theory with
$\ell=N_f-1$ massless or light, degenerate quarks $q_i$, $i=1,...,\ell$ with
mass $m_q$, and one quark $Q$ of substantial mass, $m_Q$.  The actual GMO
relation assumes SU(2) isospin symmetry and incorporates SU(3) flavor
symmetry-breaking.  For the vector meson masses squared \cite{masssquared}, the
GMO relation is \cite{gmo}
\beq 4m_{K^*}^2 = m_\rho^2 + 3m_a^2 \ ,
\label{gmo}
\eeq
where $m_\rho$ is the mass of the SU(2)-adjoint state, $\rho$, and $m_a$ is the
mass of the SU(3)-adjoint (octet) state, $\omega_8$.  Our generalization
relates the mass squared of a $Q\bar q$ vector meson, where $q$ is any of the
$\ell$ $q_i$s, to a linear combination of the squared masses of the meson that
corresponds to the operator $T_{\ell^2-1}$ of the Lie algebra of SU($\ell$) and
the meson that corresponds to the operator $T_{N_f^2-1}$ of the Lie algebra of
SU($N_f$). We denote these as $V_{\ell^2-1}$ and $V_{N_f^2-1} \equiv V_a$,
respectively. In a simple quark-model approach, one can express the mass
squared of each of these mesons as a sum of (i) a contribution $E_g$ from the
bound-state energy of the gluons, (ii) $E_{qk}$ from the bound-state energies
of the quarks, apart from their masses, and (iii) the contributions from the
hard masses of the quark and antiquark. Here we are using the quark model as a
simple method to derive the generalization of the GMO relation that encompasses
the relevant group-theoretic factors.  The energies $E_g$ and $E_{qk}$ depend
on the $n \, {}^{2S+1}L_J$ wavefunction for the state.  We leave this
wavefunction dependence implicit henceforth.  For the $Q \bar q$ meson, one
thus writes
\beq
m_{Q\bar q}^2 = \mu (E_0 + m_Q + m_q) \ ,
\label{mQqbar}
\eeq
where $\mu \sim 2\pi \Lambda$.  For $V_{\ell^2-1}$,
we have
\beq
m_{V_{\ell^2-1}}^2 = \mu (E_0 + 2m_q)
\label{mgenrho}
\eeq
The quark-model wavefunction of $V_{N_f^2-1}$ was given in
Eq. (\ref{vaexplicit}).  As was noted above, the $|V_{N_f^2-1} \rangle$ state
has a probability $1/[\ell(\ell+1)]$ of being in one of the $|q_i \bar
q_i\rangle$ states, for a total probability $1/(\ell+1)$ of being in some $|q_i
\bar q_i \rangle$ state, and a probability $\ell/(\ell+1)$ of being in the $|Q
\bar Q \rangle$ state. Hence, in the context of this quark-model approach,
\beq
m_{V_{N_f^2-1}}^2 \equiv m_a^2 = \mu \Big [ E_0 + \frac{2(m_q + \ell \,
m_Q)}{\ell+1} \Big ] \ .
\label{mgenw8}
\eeq
We now write $x \, m_{Q \bar q}^2 = y_1 \, m_{V_{\ell^2 -1}}^2 + y_2 \,
m_{V_{N_f^2-1}}^2$, where $x$, $y_1$, and $y_2$ are constants to be determined.
Without loss of generality, we can perform a rescaling to set $y_1=1$, and
after this, we set $y_2 \equiv y$.  Equating the coefficients of $m_Q$ on
either side of this equation, we find $x=2\ell/(\ell+1)$.  Equating the
coefficients of $m_q$ on either side of the equation and substituting the above
value for $x$, we obtain $y=(\ell+1)/(\ell-1)$. With these coefficients, the
$E_0$ terms are also equal on both sides of the equation. Multiplying through
by the factor $\ell+1$, we thus derive the generalization of the Gell-Mann
Okubo mass relation for mesons,
\beq
2\ell \, m_{V_{Q \bar q}}^2 =
(\ell-1) \, m_{V_{\ell^2-1}}^2 + (\ell+1) \, m_{V_{N_f^2-1}}^2 \ .
\label{ggmor}
\eeq
One readily checks that for the case of actual QCD, with $\ell=2$,
$V_{Q \bar q} = K^*$, $V_{\ell^2-1} = V_3 = \rho$, and $V_{N_f^2-1} =
\omega_8$, this relation reduces to the usual GMO relation, Eq. (\ref{gmo}).

\section{Comparison with Vector Meson Mixing}
\label{expvectormesons}

In this section we revisit $\omega$-$\phi$ mixing from the point of view of our
general-$\ell$ analysis.  It is appropriate first to recall the ways in which
our abstract analysis with $\ell=2$ (and $N_c=3$) differs from real QCD.
First, while $m_u$ and $m_d$ satisfy the criterion of being $<<
\Lambda_{QCD}$, they are not degenerate, and, second, although $m_s$ is
substantially larger than $m_u$ and $m_d$, it is not large compared to
$\Lambda_{QCD}$, indeed, $m_s/\Lambda_{QCD} \simeq 1/3$. Therefore, {\it a
priori}, one does not expect a large decoupling effect.  Indeed, one of the
most intriguing aspects of $\omega$-$\phi$ mixing is how close this is to the
decoupling limit even though $m_s/\Lambda_{QCD}$ is not $>> 1$. Moreover, in
our abstract analysis, we have turned off electroweak interactions, and the
inclusion of these, particularly electromagnetic interactions, slightly
modifies the results for general $\ell$ and, {\it a fortiori}, for $\ell=2$.

We use the conventional notation for SU(3)-singlet and SU(3)-adjoint (octet)
flavor eigenstates, which are both singlets under isospin SU(2), namely,
\beq
|\omega_0 \rangle = |(V_0)_{N_f=3} \rangle =
\Big | \frac{u \bar u + d \bar d + s \bar s}{\sqrt{3}} \Big \rangle \ ,
\label{omega0}
\eeq
and
\beq
|\omega_8\rangle = |V_8\rangle =
\bigg | \frac{u \bar u + d \bar d -2 s \bar s}{\sqrt{6}} \bigg \rangle \ .
\label{omega8}
\eeq
If the mixing were precisely as given by the decoupling result, then it would
involve a rotation of these flavor SU(3)-singlet eigenstates through the angle
mentioned in the introduction, given by
\beq
\theta_{V,dec.} = \arctan \Big ( \frac{1}{\sqrt{2}} \, \Big ) = 35.264^\circ
\quad {\rm for} \ \ell=2
\label{theta_ell2}
\eeq
to form the decoupled physical mass eigenstates
\beq
|\omega\rangle_{dec.} = |(V_0)_{\ell=2}\rangle =
\bigg | \frac{u \bar u + d \bar d}{\sqrt{2}}\bigg \rangle
\label{omega_dec}
\eeq
and $-|\phi\rangle_{dec.}$, where
\beq
|\phi\rangle_{dec.} = |V_{Q \bar Q} \rangle = |s \bar s \rangle \ .
\label{phi_dec}
\eeq
In this ideal (decoupling) mixing, Eq. (\ref{mapping_dec}) would
take the form
\beq
 \left( \begin{array}{c}
    |\omega \rangle_{dec.} \\
    -|\phi \rangle_{dec.} \end{array} \right ) =
\left( \begin{array}{cc}
     \sqrt{2/3} & 1/\sqrt{3}   \\
   -1/\sqrt{3}  & \sqrt{2/3}  \end{array} \right )
\left( \begin{array}{c}
    |\omega_0 \rangle \\
    |\omega_8 \rangle \end{array} \right ) \ .
\label{mapping_su3_dec}
\eeq
and the original mass-squared matrix in the SU(3)$_V$ flavor basis would be
\beq
(M^2)_{dec.} = \left( \begin{array}{cc}
    \frac{2m_\omega^2 + m_\phi^2}{3} &
    -\frac{\sqrt{2} \, (m_\phi^2-m_\omega^2)}{3} \\
    -\frac{\sqrt{2} \, (m_\phi^2-m_\omega^2)}{3} &
    \frac{m_\omega^2 + 2 m_\phi^2}{3} \end{array} \right ) \ .
\label{msq_decoupling_nf3}
\eeq

The actual physical angle, $\theta_{V,ph}$, by which the SU(3)-singlet
and SU(3)-octet states are rotated to form the actual physical mass eigenstates
$\omega$ and $\phi$ is close, but not exactly equal, to the decoupling value in
Eq. (\ref{theta_ell2}).  We recall the procedure for determining
$\theta_{V,ph}$, starting with the relation
\beq
 \left( \begin{array}{c}
    |\omega \rangle \\
   - |\phi   \rangle \end{array} \right ) =
R(\theta_{V,ph}) \, \left( \begin{array}{c}
                 |(V_0)_{N_f=3} \rangle \\
                 |V_8 \rangle \end{array} \right ) \ .
\label{physicalmapping}
\eeq
We next use the Gell-Mann Okubo mass relation, Eq. (\ref{gmo}), which assumes
SU(2) isospin symmetry but takes into account SU(3) breaking due to the strange
quark mass \cite{gmo}. Since SU(2) isospin symmetry is broken slightly
(explicity) by both the non-degeneracy of the hard quark masses $m_u$ and $m_d$
and, separately, by electromagnetic interactions, one expects the predictions
of this relation to be accurate to a few. For our purposes it will suffice to
use the central values of the relevant vector meson masses. From the measured
values of the neutral vector mesons $m_{K^{* 0}}=895.9$ MeV and $m_\rho=775.5$
MeV, one can solve Eq. (\ref{gmo}) for $m_a$, obtaining the result
\beq
m_a = 932.6 \ {\rm MeV}.
\label{mavalue}
\eeq
Substituting this value for $m_a$ together with the values
\beqs
m_L & = & m_\omega = 782.65 \ {\rm MeV} \cr\cr
m_H & = & m_\phi=1019.46 \ {\rm  MeV}
\label{momegaphi}
\eeqs
in the trace relation, Eq. (\ref{tracerel}), one can solve for $m_1$, with the
result
\beq
m_1 = 884.3 \ {\rm MeV} \ .
\label{m1value}
\eeq
Next, substituting this into the determinant relation Eq. (\ref{detrel}) gives
$\delta$ up to sign, namely, $|\delta| = 0.2088 \ {\rm GeV}^2$, so that
\beq
\sqrt{|\delta|} = 457.0 \ {\rm MeV}.
\label{deltavalue}
\eeq
These may be compared with the value of $|\delta|$ in the decoupling limit, as
given by Eq. (\ref{deltadec}), which is (again quoting just the central value)
$|\delta_{dec.}|=1.241 \times 10^5$ MeV$^2$, i.e.,
\beq
\sqrt{|\delta_{dec.}|} = 352.3 \ {\rm MeV} \ .
\label{deltadecsqrt}
\eeq
This decoupling value is somewhat smaller than the physical value in
Eq. (\ref{deltavalue}) and both are well below the upper bound from Eq.
(\ref{deltabound}), $|\delta| < (908 \ {\rm MeV})^2$.  In the flavor eigenbasis
(again neglecting the nonzero widths), using a minus sign for $\delta$ to agree
with Eq. (\ref{msq_decoupling_nf3}), we obtain the following form for $M^2$,
with masses quoted to the indicated accuracy:
\beq
M^2 = \left( \begin{array}{cc}
         (0.884)^2  & -(0.457)^2 \\
         -(0.457)^2 & (0.933)^2  \end{array} \right )  \quad {\rm GeV}^2 \ .
\label{msqphysical}
\eeq
Note that $m_a > m_1$, in accord with the expectation from the quark
constituency argument given above.  It is also useful to express $M^2$ as
the prefactor $m_a^2$ times a matrix of dimensionless entries to show their
relative sizes:
\beq
M^2 = (0.870 \ {\rm GeV}^2 ) \,  \left( \begin{array}{cc}
          0.899  & -0.240 \\
         -0.240 &  1  \end{array} \right ) \ .
\label{msqphysicalratios}
\eeq
Substituting these values of $m_a^2$, $m_1^2$, and $\delta$ into
Eq. (\ref{theta_dec}) yields, for the physical mixing angle,
\beq
\theta_{V,ph} \simeq 39^\circ \ ,
\label{thetaph}
\eeq
in agreement, to the requisite accuracy, with the value
$\theta_{V,ph}=38.7^\circ$ obtained in \cite{pdg} from a similar quadratic
fit and with the value $\theta_{V,ph} = (38.58 \pm 0.09)^\circ$ obtained from
a recent global fit by KLOE \cite{kloe,endep}. The fractional
deviation of this value from the decoupling value is about 10 \%:
\beq
\frac{|\theta_{V,ph}-\theta_{dec.}|}{\theta_{V,ph}} \simeq \frac{4^\circ}{
39^\circ} \simeq 0.1 \ .
\label{fractionaldev}
\eeq

In the usual analysis reviewed above, the mixing angle $\theta_V$ is
calculated in terms of physical meson masses, together with the
Gell-Mann Okubo relation embodying the breaking of SU(3) symmetry. This
analysis makes a connection with (the $\ell=2$, $N_c=3$ special case of)
our general discussion in Sect. \ref{generalmixing}, since these hadron masses
depend on $m_Q$ (with a weaker dependence on $m_u$ and $m_d$, since $m_u, \ m_d
<< \Lambda_{QCD}$).  Within the context of models such as the nonrelativistic
quark model or the MIT bag model, one has analytic formulas for hadron masses
in terms of $m_{u,d}$ and $m_s$.  At present, the most reliable determination
of hadron masses as functions of quark masses is via lattice QCD simulations.
With either approximate analytic models or dedicated lattice simulations,
one could carry out a calculation of the hadron masses in our generalized
QCD-like theory with $\ell$ light or massless degenerate quarks and one quark
of substantial mass and determine the dependence of $\theta_V$ on
$m_Q/\Lambda$.  The results would interpolate between the SU($N_f$) limit
and the asymptotic value of $\theta_V$ in Eq. (\ref{theta_dec}).

\section{Comparison with Tensor meson mixing}
\label{exptensormesons}

We next briefly review tensor meson mixing and compare it with the $\ell=2$
special case of our general-$\ell$ analysis.  The observed $J^{PC}=2^{++}$
tensor mesons $f_2(1275)$, $f_2'(1525)$, $a_2(1320)$ and $K_2^*(1430)$ form an
SU(3) $1\,^3P_2$ nonet. We consider the mixing between $f_2(1275)$ and
$f_2'(1525)$.  We use the same notation as for vector mesons with a subscript
$T$ added, so
\beq
M_T^2 = \left( \begin{array}{cc}
    m_{1_T}^2     & \delta_T \\
    \delta_T    & m_{a_T}^2    \end{array} \right ) \ .
\label{msqtensor}
\eeq
For this nonet the Gell-Mann Okubo relation reads
\beq
4m_{K_2^*}^2 = 3m_{a_T}^2+m_{a_2}^2 \ .
\label{gmotensor}
\eeq
Again, for our purposes, it will suffice to use the central values of the
measured masses.  Substituting $m_{K_2^{* 0}}=1432.4$ MeV and $m_{a_2}=1318.3$
MeV into Eq. (\ref{gmotensor}), we obtain
\beq
m_{a_T} = 1468.5 \ {\rm MeV}.
\label{matvalue}
\eeq
Inserting this together with
\beqs
m_{L_T} & = & m_{f_2} = 1275.1 \ {\rm MeV} \cr\cr
m_{H_T} & = & m_{f_2'}= 1525.0 \ {\rm  MeV}
\label{mlht}
\eeqs
in the tensor meson trace relation $m_{L_T}^2 + m_{H_T}^2= m_{1_T}^2+m_{a_T}^2$
and solving for $m_{1_T}$, we obtain
\beq
m_{1_T} = 1339.8 \ {\rm MeV} \ .
\label{m1tensorvalue}
\eeq
Next, substituting this into the tensor meson determinant relation
$m_{L_T}^2m_{H_T}^2= m_{1_T}^2m_{a_T}^2-\delta_T^2$ and solving for
$|\delta|$, we get $|\delta_T| = 0.29964 \ {\rm GeV}^2$, so that
\beq
\sqrt{|\delta|} = 457.0 \ {\rm MeV}.
\label{deltatensorvalue}
\eeq
This is slightly less than the value for complete decoupling, which, as
calculated from the tensor-meson analogue of Eq. (\ref{deltadec}), is
$\sqrt{|\delta_{T,dec.}|} = 510.9$ MeV.  In the flavor basis for the tensor
mesons, neglecting the nonzero widths, $M_T^2$ thus has the numerical form, to
the indicated accuracy,
\beqs
M_T^2 & = & \left( \begin{array}{cc}
      (1.34)^2  & -(0.547)^2 \\
     -(0.547)^2 & (1.47)^2  \end{array} \right )  \quad {\rm GeV}^2 \cr\cr
& & \cr\cr
      & = & (2.16 \ {\rm GeV}^2 ) \, \left( \begin{array}{cc}
        0.832  & -0.139 \\
       -0.139  &  1  \end{array} \right )
\label{mtsqphysical}
\eeqs

Substituting these values of $m_{a_T}^2$, $m_{1_T}^2$, and $\delta_T$ into
the tensor-meson analogue of Eq. (\ref{theta_dec}) yields, for the physical
mixing angle,
\beq
\theta_{T,ph} \simeq 29.5^\circ \ ,
\label{thetaphyst}
\eeq
in agreement, to the requisite accuracy, with the value
$\theta_{T,ph}=29.6^\circ$ obtained in \cite{pdg} from a similar quadratic
fit. The fractional deviation of this value from the decoupling
value is about 15 \%. In contrast to the vector-meson case, the physical
$f_2-f'_2$ mixing angle is thus smaller than the decoupling value;
$\theta_{T,ph} - \theta_{dec.} = 29.5^\circ - 35.26^\circ \simeq
-5.8^\circ$. As in the vector-meson case, the fact that $\theta_{T,ph}$ is
close to $\theta_{dec.}$ means that the tensor meson $f_2(1275)$ is
predominantly an isospin-singlet combination of the light quarks, $(u \bar u +
d \bar d)/\sqrt{2}$, while $f'_2(1525)$ is predominantly an $s\bar s$ state.
This is in accord with the fact that $f_2(1275)\to \pi\pi$ and $f'_2(1525)\to
K\bar K$ are experimentally the dominant decay modes for $f_2$ and $f'_2$,
respectively.

\section{Dependence of Mixing Angle in A Simple Model}
\label{thetam}

In this section we present a simple model to relate the mixing angles
$\theta_V$ and $\theta_T$ in our generalized QCD-like theory to the heavier
quark mass, $m_Q$.  Since the general results apply to both vector and tensor
meson mixing, we suppress the indices $V$ and $T$ in the notation.  We
emphasize at the outset that the actual calculation of hadron masses in QCD as
functions of the hard quark masses, in particular, $m_s$, is arguably best done
with lattice gauge theoretic methods, and the quark model that we use here is
clearly a simplification of the actual physics.  As before, we neglect
electroweak interactions.  It will be shown that the initial model that we use
is, indeed, too simplistic to describe the physics accurately, and we will then
modify it to give reasonable results.  We recall Eq. (\ref{mgenw8}) for the
mass squared of the $|V_{V_{N_f^2-1}}\rangle$ state.  By similar arguments
concerning the quark constituency of the wavefunction, we have,
\beq
m_{(V_0)_{N_f}}^2 \equiv m_1^2 = \mu \Big [ E_0 + \frac{2(\ell \,
m_q + m_Q)}{\ell+1} \Big ] \ .
\label{mgenw0}
\eeq
Thus,
\beq
{\rm Tr}(M^2) = m_H^2+m_L^2 = 2E_0 + 2(m_q+m_Q)
\label{tracerel_model}
\eeq
From the determinant condition (\ref{detrel}), we can then determine the square
of the mixing entry, $\delta^2$.  Taking $\delta$ to be negative, we thus
obtain
\beq
\delta = -\frac{2\sqrt{\ell} \,\mu \, (m_Q-m_q)}{\ell+1}
\label{deltamodel}
\eeq
and put $(M^2)_{12}=\delta$.  With these values of $m_1^2$, $m_a^2$, and
$\delta$, the diagonalization of the $M^2$ matrix would yield $\tan(2\theta) =
2\sqrt{\ell} \, (\ell-1)^{-1} \, (m_Q-m_q)/(m_Q-m_q)$. If $m_Q = m_q$, this is
proportional to 0/0, and hence indeterminate.  As discussed above, this result
is expected, since in this case there is complete SU($N_f$) symmetry, and the
$|(V_0)_{N_f}\rangle$ and $|V_{N_f^2-1}\rangle$ are degenerate in mass and may
thus be rotated by an $R(\theta)$ with any mixing angle $\theta$ to form mass
eigenstates.  However, if $m_Q > m_q$, then we may cancel the $(m_Q-m_q)$
factor through and obtain $\tan(2\theta) = 2\sqrt{\ell}/(\ell-1)$, i.e.,
$\theta = \theta_{dec.}$.  This result is unphysical, since it predicts
that there is complete decoupling of the heavier quark regardless of how small
the nonzero mass different $m_Q-m_q$ is.  This unphysical result shows that the
initial model is too simplistic. To remedy this defect, one takes account of
the fact that there is a propagator correction (for both the kinetic and mass
squared terms) in which the SU($N_f$) flavor-singlet state
$|(V_0)_{N_f}\rangle$ annihilates to an intermediate virtual purely gluonic
state and then goes back to itself again.  This annihilation process cannot
occur for the flavor-SU($N_f$) adjoint state, $|(V_0)_{N_f^2-1}\rangle$.
Modifying $m_1^2 \to m_1^2 + x_{an}$, where
$x_{an}$ denotes the annihilation contribution to the squared mass, we then
have, in this quark-model approach,
\begin{widetext}
\beq
\mu^{-1} M^2 = \left( \begin{array}{cc}
 E_0 + \frac{2(\ell \, m_q + m_Q)}{\ell+1} + x_{an}  &
 -\frac{2\sqrt{\ell} \, (m_Q-m_q)}{\ell+1} \\
 -\frac{2\sqrt{\ell} \, (m_Q-m_q)}{\ell+1} &
 E_0 + \frac{2(m_q + \ell \, m_Q)}{\ell+1} \end{array} \right ) \ .
\label{msqqm}
\eeq
\end{widetext}
The diagonalization of this matrix yields
\beq
\tan(2\theta) = \frac{2\sqrt{\ell}}{\ell-1-\xi}
\label{tan2theta_modified}
\eeq
where
\beq
\xi \equiv \frac{(\ell+1)x_{an}}{2(m_Q-m_q)} \ .
\label{xi}
\eeq
Provided that $x_{an} \ne 0$, it follows that (i) as $(m_Q - m_q)/x_{an} \to
0$, $\xi \to \infty$ and hence $\tan(2\theta) \to 0$, so that $\theta \to 0$ or
$\pi/2$, and (ii) as $(m_Q -m_q)/x_{an} \to \infty$, $\xi \to 0$, and $\theta
\to \theta_{dec.}$.  This prediction makes physical sense and agrees with our
general result in Eq. (\ref{theta_dec}). This analysis, by itself, does not
determine the sign of $x_{an}$.  If $x_{an} < 0$, then $\theta$ increases
monotonically from 0 to $\theta_{dec.}$ as $(m_Q-m_q)/x_{an}$ increases from 0
to $\infty$. If $x_{an} > 0$ and $\theta=\pi/2$ for $m_Q-m_q \to 0^+$, then as
$(m_Q-m_q)/x_{an}$ increases from $0^+$ to $\infty$, $\theta$ decreases
monotonically from $\pi/2$ to $\theta_{dec.}$, passing through $\pi/4$ as
$\xi$ increases through the value $\ell-1$.

\section{Other Meson Mixings}
\label{axialvectormesons}

Similar mixings of SU(3)-singlet and SU(3)-octet states occur for other mesons.
As is well-known, the mixing of the $J^{PC}=1^{-+}$ pseudoscalar mesons $\eta$
and $\eta'$ from the $1 \, {}^1S_0$ nonet is complicated by several effects,
including the fact that the $\eta$ is an approximate Nambu-Goldstone boson
(NGB), and the fact that there is a splitting between the octet of approximate
Nambu-Goldstone bosons and the $\eta'$ due to the breaking of the
flavor-singlet axial vector global symmetry by instantons, which has the
consequence that the $\eta'$ is not an approximate NGB \cite{etaetaprimeglue}.
An additional effect involves mixing of the $\eta$ and $\eta'$ with the
$0^{-+}$ glueball.  Since we have focused here on mixing in a generalized
QCD-like theory and its connection with decoupling, the illustration with
vector and tensor mesons is sufficient for our purposes.

A comment on axial vector meson mixing is also in order here. In the
quark model, two nonets of $J^P=1^+$ axial-vector mesons are expected as
orbitally excited quark-antiquark bound states. In the usual spectroscopic
notation, these are $1 \, {}^3P_1$ and $1 \, {}^1P_1$. These two nonets have
different $C$ quantum numbers for their respective neutral mesons, namely $C=+$
and $C=-$.  Experimentally, the $1 \, {}^3P_1$ nonet consists of $a_1(1260)$,
$f_1(1285)$, $f_1(1420)$ and $K_{1A}$, while the $1 \, {}^1P_1$ nonet contains
$b_1(1235)$, $h_1(1170)$, $h_1(1380)$ and $K_{1B}$.  The non-strange axial
vector mesons, for example, the neutral $a_1(1260)$ and $b_1(1235)$ cannot mix
because of their opposite $C$-parities. In contrast, $K_{1A}$ and $K_{1B}$ do
mix to form corresponding physical mass eigenstates.  This complicates the
analysis of the mixings of the SU(3)-singlet and SU(3)-octet mesons in the $1
\, {}^3P_1$ and $1 \, {}^1P_1$ nonets.  For recent discussions of these
mixings, see \cite{axialvectormesons} and references therein.

\section{Conclusions}

In this paper we have discussed the mixing of quark flavor-eigenstates to form
low-lying vector and tensor meson mass eigenstates in a QCD-like theory with
$\ell$ massless or light, degenerate quarks and one quark of substantial mass,
$m_Q$.  We have derived the asymptotic value of the mixing angle in the limit
as $m_Q/\Lambda$ gets large.  We have also presented a generalization of the
Gell-Mann Okubo mass relation for this theory.  Finally, we have remarked on
how the $\ell=2$ special case of our results relate to the observed
$\omega$-$\phi$ and $f_2(1275)$-$f'_2(1525)$ mixings.

\begin{acknowledgments}

This research was partially supported by the R.O.C. National Science Council
Grant NSC-100-2112-M-001-009-MY3 (H.-Y. C.) and the U.S. National Science
Foundation grant NSF-PHY-09-69739.

\end{acknowledgments}


\begin{thebibliography}{99}
\bibitem{cpv}
%
In our discussion of meson mixing, we rely on the empirical fact that QCD
conserves P, C, and CP. This is automatic if $m_q=0$, since then the
CP-violating angle $\bar\theta_{CPV} = \theta_{CPV} - {\rm arg}({\rm det}(M_u))
- {\rm arg}({\rm det}(M_d))$ can be rotated away, where $\theta_{CPV}$ is the
coefficient of the instanton-induced topological term $g_s^2(32\pi^2)^{-1}
G^a_{\mu\nu}\tilde G^{a \mu\nu}$ in the QCD Lagrangian. For $m_q \ne 0$, we
implicitly assume a solution of the strong CP problem.
\bibitem{vacalign}
%
Although electromagnetic interactions are a small perturbation, their
presence, together with vacuum alignment arguments, implies the absence of
charge-violating quark condensates such as $\langle \bar u d \rangle$,
$\langle \bar u s \rangle$, etc.
\bibitem{masssquared}
%
Here and below, e.g., when using the Gell-Mann Okubo mass relation and giving
various quark model formulas, we follow the common practice of using the
squared masses of the mesons rather than the masses themselves, because for
bosons it is the squared masses that appear in effective Lagrangians.
\end{thebibliography}
\end{document}